\documentclass[11pt]{article}
\pdfoutput=1
\usepackage{jheppub}
\usepackage{subfig}

\newcommand\beq{\begin{eqnarray}}
\newcommand\eeq{\end{eqnarray}}
\newcommand\be{\begin{equation}}
\newcommand\ee{\end{equation}}

\title{Holographic Black Hole Chemistry}

\preprint{\today}

\author{Andreas Karch and Brandon Robinson}
\affiliation{Department of Physics, University of Washington, Seattle, Wa, 98195-1560, USA}
\emailAdd{akarch@uw.edu, robinb22@uw.edu}

\abstract{Thermodynamic quantities associated with black holes in Anti-de Sitter space obey an interesting identity when the cosmological constant is included as one of the dynamical variables, the generalized Smarr relation. We show that this relation can easily be understood from the point of view of the dual holographic field theory. It amounts to the simple statement that the extensive thermodynamic quantities of a large $N$ gauge theory only depend on the number of colors, $N$, via an overall factor of $N^2$.
}

\begin{document}
\maketitle
%\tableofcontents

%%%%%%%%%%%%%%%%%

\section{Introduction}

The thermodynamics of black holes in Anti-de Sitter (AdS) space has a long history, starting with the pioneering work of Hawking and Page \cite{Hawking:1982dh}. This subject has found a solid theoretical foundation in terms of the holographic duality, where the thermal properties of AdS black holes can be reinterpreted as those of a conformal field theory at finite temperature \cite{Witten:1998zw}. The result is that the grand canonical free energy $\Omega$ is expressible in terms of the on-shell action $S_{os}$ of the Euclidean bulk solution as
\beq
\label{omega}
T S_{os} = \Omega(\mu,T) = U - TS - \mu Q
\eeq
where $U$, $T$, $S$, $\mu$ and $Q$ denote the energy, temperature, entropy, chemical potential and charge of the black hole. For black branes with a homogeneous planar horizon the extensive quantities $\Omega$, $U$, $S$ and $Q$ are spatially independent. So their entire dependence on the volume $V$ can be expressed in terms of densities:
\beq \Omega = V \omega, \quad U = V u, \quad S = V s, \quad Q = V q. \eeq
As a consequence the pressure, $p= - d\Omega/dV|_{T,\mu} = - \omega $, is simply minus the grand canonical free energy density. For spherical horizons, the spatial volume of the dual field is finite and completely determined by the radius $R$ of the sphere on which the field theory lives.  In this case, the free energy depends on $R$ explicitly via dimensionless ratios like $TR$ and $\mu R$ in addition to the overall prefactor of $V = s_n R^n$, where $n$ is the number of spatial dimensions of the field theory and $s_n$ is the volume of the unit $n$-sphere. The most reasonable definition of pressure in this case seems to be\footnote{That is, we vary the volume of the system by changing the curvature radius of the sphere the field theory lives on. One could try to divorce the volume of the system from the curvature radius of the background geometry by putting the fluid into a box of size much less than $R$ whose volume $V$ can be independently varied. But as long as the microscopic scales in the system, like $T$ and $\mu$, are of the same order of $R$, we would get a strong sensitivity on the shape and details of the box, which seems undesirable. If the microscopic scales are much smaller than the size of the sphere, this is of course not an issue. But in this case the second term in \eqref{pressure} is also negligible to begin with; one is back to an approximately flat space.}
\beq
\label{pressure}
p = - \frac{dR}{dV} \frac{d \Omega}{dR} =  - \frac{R d_R \Omega}{n V} = - \omega - \frac{R d_R \omega}{n} \equiv
-\omega +p_R/n.
\eeq
In either case, the field theory pressure is completely determined once the full function form of $\Omega$ is known. For the case of hyperbolic horizons, even though the spatial volume for the dual field theory is non-compact, the finite length scale associated with the background geometry gives rise to an effective pressure by the same analysis.

A different notion of pressure for black hole thermodynamics has recently been discussed \cite{Kastor:2010gq, Dolan:2012jh}. It was noted that the standard first law of black hole thermodynamics relating changes in the mass $M$ and the area $A$ as
\beq dM = \frac{\kappa}{8 \pi} dA \eeq
seems to miss the pressure-volume term when compared to the standard first law, $dU=TdS-p_bdV_b$. One can recover a $p_bV_b$ term  \cite{Caldarelli:1999xj,Creighton:1995au,Gibbons:2004ai,Kastor:2009wy,Cvetic:2010jb} by identifying the cosmological constant as a pressure term $p_b \equiv - \Lambda/(8 \pi)$ and the `thermodynamic volume' as the corresponding conjugate variable, $V_b \equiv \partial M/\partial p_b |_A$. We use the subscript $b$ for ``bulk" in order to remind the reader that these quantities are not pressure and volume of the holographically dual quantum field theory.
 Studying the generalized black hole thermodynamics including this new term has been dubbed ``black hole chemistry" in \cite{Kubiznak:2014zwa} and we will continue to use this term despite the slight abuse of language.

 As we reviewed above, the pressure of the dual quantum fluid is completely determined by standard thermodynamic relations once the on-shell action has been identified as the grand canonical free energy. Thus, it is not a quantity that should nor can be defined separately.  However, from the bulk point of view the metric's non-trivial dependence on the radial coordinate means that space is not homogeneous. It then, from the higher dimensional perspective, indeed makes sense to identify the cosmological constant as a pressure and $V_b$ as an effective bulk (thermodynamic) volume of the black hole. Note that the notion of thermodynamic volume can differ from the geometric volume that would be calculated from simply integrating the volume form from the origin to the outer horizon of the black hole \cite{Cvetic:2010jb}.

In black hole chemistry, one finds a remarkable identify obeyed by the thermodynamic quantities:
\beq
\label{smarr}
(d-3) M = (d-2) T S - 2 P_b V_b + (d-3) \mu Q,
\eeq
where $d=n+2$ is the number of spacetime dimensions of the gravitational theory.
This Smarr relation generalizes a similar identity that had first been derived for flat space black holes in $d=4$ \cite{Smarr:1972kt} and can easily be extended to encompass higher dimensional spinning black holes as well \cite{Myers:1986un}.
In this note we'd like to give a holographic identification of $p_b$ and $V_b$ as well as an holographic derivation of the generalized Smarr formula \eqref{smarr}.

\section{Holographic Smarr Relations}

\subsection{A universal relation in large $N$ field theories}

Since the pressure of the holographic fluid is already fixed by the on-shell action, the bulk cosmological constant, which yields $p_b$, can clearly not play the role of pressure on the field theory side. As has already been emphasized in earlier work on holographic black hole chemistry \cite{Kastor:2014dra, Johnson:2014yja,Zhang:2014uoa,Zhang:2015ova,Caceres:2015vsa}, varying the cosmological constant $\Lambda$ in the bulk essentially amounts to changing the number of colors in the field theory. Unlike changes in physical properties like the temperature and the chemical potential, the number of colors, $N$, is not a standard thermodynamical variable. But one can ask the same questions also on the field theory side: how does the free energy of the system change as we vary the number of colors? We can define a color susceptibility $\chi_{N^2}$ as
\beq
\chi_{N^2} = \left . \frac{\partial \Omega}{\partial N^2} \right |_{\lambda,\mu,T,R},
\eeq
where as usual in holography, we are working at fixed `t Hooft couling, $\lambda$.  Unless the following discussion warrants clarification, we will omit explicit reference to $\lambda$. Generically $\Omega$ will be a highly non-trivial function of $N$.  That is, the dynamics of the field theory crucially depends on the number of colors. But in the limit of a large number of colors, $N$ only shows up as an overall prefactor in the grand canonical free energy:
\beq \Omega(N, \mu,T,R) = N^2 \Omega_0( \mu,T,R). \eeq
Correspondingly $\chi_{N^2}$ obeys the trivial relation
\beq \label{holosmarr} N^2 \chi_{N^2} = \Omega. \eeq
As we will soon see, this simple relation will turn out to be the holographic origin of the Smarr relation \eqref{smarr}. Note that this holographic Smarr relation \eqref{holosmarr} is completely universal: It is true in any large $N$ gauge theory irrespective of the details of the equation of state. It applies equally well to conformal theories, confining theories, or those with unusual scalings such as hyperscaling violating theories. To be clear, the interpretation here is that $N^2$ is holographically standing in for the AdS length scale, $L$, in Planck units, i.e. $L^{d-2}/G_N$ where $G_N$ is Newton's constant. This quantity determines up to numerical factors the central charge of the dual field theory. In the examples we are considering here based on the worldvolume theory of Dd branes this central charge is proportional to $N^2$ as appropriate for gauge theories. However, there exist holographic theories in which $L^{d-2}/G_N$ corresponds to a different power of $N$ in the dual theory. For example, M2 brane theories famously have $N^{3/2}$ scaling, and M5 branes have $N^3$ scaling. In all of these cases, a similarly trivial analog of \eqref{holosmarr} can be formulated by directly varying with respect to the central charge instead of first reexpressing it in terms of $N$.

Given the universal Smarr relation \eqref{holosmarr}, many derived identities can be formulated for a particular system using the equation of state relating $p_R=-R \partial_R \omega$, $s=-\partial_T \omega$, and $q=-\partial_{\mu} \omega$, which implicitly appear on the right hand side via $\Omega=U-TS-\mu Q$. Of particular interest are conformal field theories, whose equation of state is fixed by considering the behavior of the thermodynamic quantities under an infinitesimal scale transformation (under which all energies are rescaled by $\lambda=1+d\lambda$ and all length by $\lambda^{-1}=1-d\lambda$):
\beq
d U =  U \, d\lambda, \quad dS = 0 , \quad d Q =  0 , \quad  dV = - n V \, d\lambda.
\eeq
The fact that a scale transformation takes one physical configuration into another, and so has to represent a set of variations consistent with the first law of thermodynamics
\beq
dU = T dS - p dV + \mu dQ,
\eeq
immediately yields the equation of state
\beq
\label{eos}
U = n p V.
 \eeq
Using the definition of the pressure from \eqref{pressure},  \eqref{eos} can be equivalently written as
\beq
\label{eostwo}
(n+1) u = n T s + n \mu Q - R \partial_R \omega.
\eeq
What we will show in the next subsection is that the universal Smarr relation \eqref{holosmarr} together with the conformal equation of state \eqref{eos} implies the bulk Smarr relation \eqref{smarr}. For holographic spacetimes dual to non-coformal, large $N$ gauge theories the bulk Smarr relation hence will have to be modified together with the equation of state, but the universal relation \eqref{holosmarr} will remain valid.

\subsection{Holographic Dictionary}

In order to recover the bulk Smarr relation, we need to carefully relate bulk and field theory quantities. Employing the standard lore that bulk thermodynamic quantities are simply equated with their boundary analogs is a little too careless. Using the standard thermodynamic definitions on both sides, one has to account for the factors of the curvature radius, $L$, appearing between the two sets of variables.  These factors of $L$ become crucial when one is studying the response of the theory to variations in $\Lambda=(d-1)(d-2)/(2 L^2)$.

From the perspective of the dual field theory, the most important effect of varying $L$ is, as advertised, a change in $N$. The basic holographic dictionary identifies
\beq
\alpha \frac{L^{d-2}}{16 \pi G_N} =  N^2.
\eeq
The purely numerical factor, $\alpha$, depends on the details of the particular holographic system being studied, but it drops out from the product $p_b V_b$.  Thus for our purposes in considering the Smarr relation, $\alpha$ is irrelevant. We defined previously
\beq
\label{defpress}
p_b \equiv - \Lambda/(8 \pi), \quad \quad V_b \equiv \partial M/\partial p_b |_{A,Q_b} \eeq
where again the subscript $b$ refers to a bulk quantity.
For functions that only depend on $L$ through $N^2$, we would have
\beq
\label{partialone}
-2 \Lambda \partial_{\Lambda} = L \partial_L = (d-2) N^2 \partial_{N^2},
\eeq
which shows that $\alpha$ does not enter into $p_bV_b$.  In the last equality in \eqref{partialone}, it is important to take the partial derivatives at fixed $G_N$. This is how the calculation proceeds in standard black hole chemistry \cite{Kubiznak:2014zwa}, but it is a requirement that we will soon relax.

The relation between $\Lambda$ and $N^2$ however is not the only place $L$ appears in the dictionary. In addition to changing $N$, variations in $L$ will also vary the curvature radius $R$ of the manifold on which our field theory is formulated. Note that in standard black hole thermodynamics, we write the metric of a generic homogeneous black hole as
\beq\label{BHmetric}
ds^2_b = - h(r) dt^2 + dr^2/h(r) + r^2 d \Sigma_k^2,
\eeq
where $d\Sigma_k^2$ is the dimensionless metric on a unit sphere, plane or hyperboloid corresponding to $k=1,\, 0,\, -1$ respectively. For an asymptotically AdS$_d$ space, the blackening function at large $r$ approaches
\beq
h(r) = \frac{r^2}{L^2} + \ldots .
\eeq
To read off the field theory metric, we need to multiply \eqref{BHmetric} with an overall factor of a defining function with a double zero at $r=\infty$ and then evaluate the metric at $r=\infty$ \cite{Witten:1998qj}. Choosing the defining function\footnote{Alternatively, we could use $r^{-2}$ or $\tilde{R}^2/r^2$ as a defining function with $\tilde{R}$ being an arbitrary length scale. For the former choice, the metric would read as $ds^2 = -L^2dt^2 +d\vec{x}^2$. In the latter case, $\tilde{R}^2$ would appear in front of the spatial metric and $dt^2$ would come with a factor of $\tilde{R}^2/L^2$. Of course any choice would give physically equivalent results. We find it easier to work with the dimensionless defining function $L^2/r^2$, so that all coordinates retain their bulk dimensionalities and we do not need to introduce yet another separate scale $\tilde{R}$.} to be $L^2/r^2$, we see that the boundary metric is
\beq\label{bdrymetric}
ds^2 = -dt^2 + L^2 d\Sigma_k^2. \eeq
That is, the bulk curvature radius $L$ in the field theory double features as playing the role of $N$, the number of colors, as well as the curvature radius $R=L$ of the spatial metric. This is even true in the planar, $k=0$, case. Note that, even for planar geometry, we chose to use dimensionless coordinates in $d\Sigma_k^2$, and so $L$ sets the overall length scale in the field theory. In particular, the field theory volume $V$ scales as $R^n=L^n$. That is, the relation \eqref{partialone} between partial derivatives, when relating bulk to boundary quantities, really has to be taken to be
\beq
\label{partialtwo}
-2 \Lambda \partial_{\Lambda} = L \partial_L = (d-2) N^2 \partial_{N^2} + R \partial_R.
\eeq
The notion of bulk volume and pressure as defined in the black hole chemistry literature mixes up the notion of boundary pressure and volume (with their roles reversed) with variations of $N$.

There is one more power of $L$ hiding in the standard gravitational definitions of the thermodynamic quantities. In the study of black holes and black hole chemistry one usually works with an action
    \beq
    S =  \frac{1}{16 \pi G_N} \, \int d^dx \, \sqrt{-g} \, \left (R - 2 \Lambda - F_b^2 \right ) .
    \eeq
    Note that the standard convention is to pull out an overall factor of $G_N$, and so Einstein-Maxwell theory is commonly written in terms of a bulk field strength of dimension 1. To identify the leading (constant) term in the corresponding gauge potential, we need to convert to a canonically normalized field strength of dimension 2 via
    \beq
    A_b = L A, \quad \quad \mu_b = L \mu, \quad \quad Q_b = L^{-1} Q .
    \eeq
  This has interesting consequences. The variation defining the bulk ``volume" in black hole chemistry via \eqref{defpress} is entirely defined in terms of bulk quantities. Thus, the variation is done at fixed $Q_b$. To compare to the field theory thermodynamics, we have to do all variations at fixed $Q$. This can be accomplished by carefully tracking the $L$ dependences,
  \beq
  \label{partialq}
  \left . \partial_L f(L,Q(Q_b,L)) \right |_{Q_b} = \left . \partial_L f \right |_{Q}  + \left . \partial_Q f \right |_{L}  \left . \partial_L Q  \right |_{Q_b} =
  \left . \partial_L f \right |_{Q}  + \frac{Q}{L} \left .  \partial_Q f \right |_{L},
  \eeq
  for any function $f$. This is simply stating that a variation $dL$ of $L$ that leaves $Q/L$ fixed needs to be accompanied with a variation in $Q$ with $dQ=dL$.

The dictionary is now completely fixed. In terms of blackening function and bulk gauge field
\beq\label{AdSRN}
h(r)=\frac{r^2}{L^2}+k-\frac{m}{r^{d-3}} + \frac{q^2}{r^{2d-6}}, \quad  A_b = \left ( - \frac{1}{c} \frac{q}{r^{d-3}} + \mu_b \right ) dt, \quad c = \sqrt{\frac{2(d-3)}{d-2}},
\eeq
we can read off the field theory thermodynamic quantities  \cite{Chamblin:1999tk}. The intensive variables are
\beq
T=\frac{(d-1) r_+^2 + (d-3) L^2 (k-c^2 \mu_b^2)}{4 \pi L^2 r_+},
\quad
\mu =\frac{\mu_b}{L} = \frac{q}{cLr_+^{d-3}}, \quad R=L \eeq
whereas the extensive quantities are
\begin{eqnarray}\label{AdSRNExt}
\nonumber Q&=& L Q_b = \sqrt{2(d-2)(d-3)} \frac{L s_{d-2} q}{8 \pi G_N}, \\
\nonumber  U &=&M= \frac{(d-2) s_{d-2} m}{16 \pi G_N}, \\
S  &=& \frac{A}{4G_N} = \frac{ s_{d-2} r_+^{d-2}}{4 G_N}.
\end{eqnarray}
The horizon radius, $r_+$, is the largest real positive root of $h(r)$. Note that all extensive quantities come with a prefactor of $(16 \pi G_N)^{-1}$. This is natural as they derive from variations of the on-shell action with respect to an intensive variable. $G_N$ does not enter the solution itself but rather the action evaluated on the solution.

\subsection{Holographic derivation of the Smarr relation}

What we would like to show is that the somewhat mysterious Smarr relation \eqref{smarr}, which can be derived in the bulk from scaling considerations, can naturally be understood as a consequence of processing the universal holographic Smarr relation \eqref{holosmarr} that expresses the simple large $N$ scaling of the free energy density with the conformal equation of state \eqref{eos}. In the following section, we will confirm that more general Smarr-like relations can be derived from the universal form for different asymptotic geometries by modifying the equation of state. But for now, let us work with the conformal case to reproduce the known form of the Smarr formula. From the definition of $p_b$ and $V_b$ we have
\beq
\label{workwith}
-2 p_b V_b = - \left . 2 \Lambda \partial_{\Lambda} M \right |_{S,Q_b} =
 \left. (d-2) N^2 \partial_{N^2} U \right |_{S,Q} + \left . R \partial_R U \right |_{S,Q} + \left . Q \partial_Q U,
 \right |_{S,L},
\eeq
where we used the expression of $L$ derivatives in terms of boundary quantities from \eqref{partialtwo} as well as the relation between derivatives at fixed $Q$ versus fixed $Q_b$, \eqref{partialq}. Evaluating each of the three terms individually, we note that since $U$ is the Legendre transform of $\Omega$, see \eqref{omega}, we have for the first term in \eqref{workwith}
\beq
   \left . N^2 \partial_{N^2} U \right |_{S,Q} = \left . N^2 \partial_{N^2} \Omega \right |_{\mu,T} = \Omega.
\eeq
In the last step, we used the universal Smarr relation \eqref{holosmarr}. The derivative in the third term in \eqref{workwith} is the defining relation for the chemical potential $\mu$, and so it simply evaluates to $\mu Q$. For the second term in \eqref{workwith}, we use the definition of pressure from \eqref{pressure} and again use the fact that $U$ and $\Omega$ are related by Legendre transforms\footnote{Explicitly, take
\begin{eqnarray} \partial_R U |_{S,Q} &=& \partial_R
\left . \left ( \Omega(\mu(S,Q,R),T(S,Q,R),R) + \mu Q + S T \right ) \right |_{S,Q} \\ \nonumber
&=& \left . \partial_{\mu} \Omega \right |_{T,R} \left . \partial_R \mu \right |_{S,Q} +
 \left . \partial_{T} \Omega \right |_{\mu,R} \left . \partial_R T \right |_{S,Q}
+  \left . \partial_{R} \Omega \right |_{\mu,T} + Q \left . \partial_{R} \mu \right |_{S,Q} +S \left . \partial_{R} T \right |_{SQ} =  \left . \partial_{R} \Omega \right |_{\mu,T}
\end{eqnarray}
where we used $\left . \partial_{\mu} \Omega \right |_{T,R}=-Q$ and $\left . \partial_{T} \Omega \right |_{\mu,R}=-S$.
}
to arrive at,
\beq
\left . R \partial_R U \right |_{S,Q} = \left . R \partial_R \Omega \right |_{\mu,T} = -n p V = - U.
\eeq
In the last step we used the conformal equation of state \eqref{eos}. Putting the three terms back together we finally arrive at
\beq
- 2 p_b V_b = (d-2) \Omega + \mu Q - U = (d-3) U - (d-2) T S - (d-3) \mu Q,
\eeq
which is, in fact, exactly the standard Smarr relation \eqref{smarr}. Note that, as expected, we needed to use both the universal formula \eqref{holosmarr} as well as the conformal equation of state \eqref{eos} to derive this result.

\subsection{Extracting the Boundary pressure from the Bulk}

As we have seen, one must be careful in extracting boundary thermodynamic quantities from the bulk, and this is doubly true for pressure.  The notion of a bulk volume one obtains from varying the bulk curvature radius $L$ from the field theory point of view mixes up two completely separate notions. Varying $L$ induces a variation of $R$, the length scale governing the field theory volume and curvature. This variation gives a contribution to the bulk {\it volume} which actually corresponds to the {\it pressure} of the holographically dual field theory. In addition, varying $L$ also varies the number of colors, $N$. This is a very different notion in the holographic dual as it takes us from one theory to another. As we have shown above, the corresponding response in a large $N$ field theory is essentially trivial, a fact that allowed us to derive the bulk Smarr formula from general field theory considerations.

From the point of view of holography, it would be desirable to already in the bulk disentangle the notion of changing the volume from the notion of changing the number of colors.
This will be crucial below when we consider explicit examples. In order to compare the equation of state \eqref{eostwo} to an expression derived from bulk quantities, we also need to be able to calculate $R \partial_R$ at fixed $N$ in the bulk.
The easiest way to do this is to vary both the AdS length, $L$, and $G_N$ (and hence the Planck length) simultaneously such that $N^2$ is unchanged.  In the bulk, this plays out by noting the standard holographic relation
\be
\frac{L^{d-2}}{G_N}\sim N^2.
\ee
In all of the gravitational formulae for thermodynamic quantities, we can make this replacement and then carry out the variations with respect to $L$ at fixed $N^2$ straightforwardly. On the other hand, if we wish to only vary $N$ without varying $R$, and hence the field theory volume, we can vary just $G_N$ at fixed $L$. To summarize, for any function $f$ we have the following dictionary
\beq
\mbox{Changing } N: \, \partial_{N^2} f |_R = \partial_{G_N^{-1}} f |_L, \quad \quad
\mbox{Changing } R: \, \partial_R f  |_{N^2} = \partial_L f |_{L^3/G_N}.
\eeq
In contrast throughout the black hole chemistry literature, the relevant variation that appears, $\partial_L f|_{G_N}$, corresponds to changing both $N$ and $R$.

\subsection{Applications}
In order to demonstrate the utility of this approach and explicitly check its validity, we will take a tour of prototypical examples. Beginning with a $d$ dimensional AdS Reissner-Nordstrom black hole
\be
ds^2 = -h(r)dt^2+\frac{dr^2}{h(r)}+r^2d\Omega_{d-2}^2,
\ee
where the blackening function $h$ and gauge field are given in \eqref{AdSRN}. It is well documented in the black hole chemistry literature, that the standard bulk Smarr relation \eqref{smarr} holds for this black hole. What we would like to confirm is that this black hole also satisfies the equation of state \eqref{eostwo} that we derived above.  Using the form of the extensive quantities in \eqref{AdSRNExt} and the relation
\be
m=k r_+^{d-3}+\frac{r_+^{d-1}}{L^2}+\frac{q^2}{r_+^{d-3}},
\ee
we can easily read off the equation of state by comparing $TS+\mu_bQ_b$ to the ADM mass $M$, which reads
\be\label{bulkeos}
(d-1)M = (d-2)(TS+\mu_b Q_b) + 2 k\frac{s_{d-2}}{16\pi G}r_+^{d-3}.
\ee
How does this map onto \eqref{eostwo}? The left hand side and the first two terms on the right hand side match \eqref{eostwo} with $n=d-2$ but identifying the last term as $p_R=-R \partial_R \omega$ takes more work. From the discussion in the previous subsection, we can find $p_R$ by first calculating the thermodynamic potential
\beq \Omega = E- TS - \mu Q = -\frac{s_{d-2}}{16 \pi G_N} \left ( \frac{r_+^{d-1}}{L^2} + \frac{q^2}{r_+^{d-3}} - k r_+^{d-3} \right ),
\eeq
and then taking derivatives with respect to $L$ at fixed $L^{d-2}/G_N$, $\mu$ and $T$. In order to do so, we need to convert all of the variables ($r_+,\,q$) in \eqref{bulkeos} to quantities that are held fixed.  That is we use $G_N \sim L^{d-2} N^{-2}$, $r_+ \sim L^2 T$ and $q \sim \mu Lr_+^{d-3}~$.
With these replacements the first and second term in the density $\omega=\Omega/(s_{d-2} R^{d-2})$ scale as $L^0$, whereas the last term proportional to $k$ scales as $L^{-2}$. Consequently
\be
\left . R \partial_R \omega \right |_{N,\mu,T} = \left . L\partial_L \omega \right |_{L^{d-2}/G_N,\mu,T} = -2\frac{k r_+^{d-3}}{16 \pi G_N L^{d-2}}.
\ee
This is exactly what is needed to confirm that \eqref{eostwo} indeed holds for these black holes.

As a non-trivial check of our construction, we can turn our focus to holographic systems with a different equation of state. A simple class of examples are geometries dual to large $N$ gauge theories with hyperscaling violation.  Top-down examples of such theories are maximally supersymmetric gauge theories in $n+1$ dimensions dual to black Dn branes, which are described by \cite{Itzhaki:1998dd}
\beq\label{dpdqgeometry}
ds^2&=& H^{-\frac{1}{2}}\left(-h(r) dt^2+dx_n^2\right) + H^{\frac{1}{2}}\left(\frac{dr^2}{h(r)}+r^2d\Omega_{8-n}^2\right),\\\nonumber
e^\Phi &=& H^{\frac{3-n}{4}},\qquad C_{01\ldots n} = H^{-1},
\eeq
where $H=1+ \left(\frac{L}{r}\right)^{7-n}$ and $h(r) = 1-\left(\frac{r_+}{r}\right)^{7-n}$.  The temperature can be found by calculating the surface gravity $\kappa =2\pi T$ or demanding that the analytic continuation of the time direction be free of conical singularities such that
\be
T = \frac{7-n}{4\pi L}\left(\frac{r_+}{L}\right)^{\frac{5-n}{2}},
\ee
which demonstrates the scaling behavior of the horizon radius $r_+ \sim T^{\frac{2}{5-n}}L^{\frac{7-n}{5-n}}$ as seen above for the $n=3$ case.  In $n+2$ dimensional Einstein frame with $L=1$, the metric takes the form
\be
ds_{n+2}^2 = r^{2(8-n)}H^{\frac{1}{n}}\left(-h(r) dt^2 +dx_n^2+\frac{dr^2}{h(r)}\right).
\ee
From this, we can read off the scaling for the field theory entropy, energy, and charge (if present) densities \cite{Dong:2012se}
\be
[s]=n-\theta,\quad [u] = n+1-\theta, \quad [q] = n-\theta,\quad \theta=-\frac{(n-3)^2}{5-n}.
\ee
Again using these scaling relations in a first law calculation, we find that the equation of state is given by
\be\label{dpdqeos}
(n+1-\theta)u =(n-\theta)(Ts+\mu q) = (n+1-\theta) n p .
\ee
These identities are indeed obeyed by the thermodynamic quantities derived in \cite{Itzhaki:1998dd} as can e.g.  be seen from the detailed expressions presented in \cite{Karch:2009eb}. For the non-hyperscaling violating case, i.e. $n=3$ or $\theta=0$, we recover the equation of state found previously. $\omega$ still scales as $N^2$ as appropriate for a large $N$ gauge theory and so the `universal' holographic Smarr relation, \eqref{holosmarr} still holds for any $n$. So direct analogs of both of the relations we used in the conformal case, the equation of state \eqref{eostwo} as well as the universal Smarr formula \eqref{holosmarr} still hold for the general Dn brane. As expected, only the equation of state is modified due to the presence of the non-trivial hyperscaling violating exponent.

\section{Finite $N$ corrections}

Much of our analysis was based on the fact that, in the large $N$ limit, all extensive thermodynamics quantities in a gauge theory simply scale with an overall prefactor of $N^2$. This allowed us to derive the universal Smarr relation \eqref{holosmarr} from which all other Smarr formulae followed using the equations of state. Note that this simple fact also has far-reaching consequences when thinking about the phase diagram of the theory. For the free energy
\beq \Omega(N,\mu,T) = N^2 \Omega_0(\mu,T) \eeq
to have any non-trivial phase transitions, these discontinuities must all arise from the behavior of $\Omega_0$. No non-trivial phase transitions can possibly occur as a function of $N$. That is, adding the cosmological constant $\Lambda$ as a new thermodynamical parameter may give the phase diagram an extra dimension. But when properly organized into $N$, $\mu$ and $T$, we see that the phase diagram is just a trivial extension of the phase diagram living in the $\mu$-$T$ plane. In fact, the black hole chemistry literature indeed finds \cite{Kubiznak:2012wp} that the standard first order Hawking phase transition simply extends into a line of first order phase transitions when $\Lambda$ is included as a parameter, while the Reisner-Nordstrom black hole retains its Van der Waals behavior that was already found in the traditional holographic analysis of the same system in \cite{Chamblin:1999tk}. As we have seen here, this is  a trivial consequence of the simple $N$ scaling of the free energies. The fact that the phase diagrams in \cite{Kubiznak:2012wp} look slightly non-trivial is simply due to the fact that, as we showed in here, varying with respect to $\Lambda$ mixes up the trivial variation of $N$ with a variation of the volume the field theory lives on.

The situation changes dramatically when we go beyond the leading large $N$ limit. On the bulk side, this corresponds to including higher curvature terms. In fact, studies of black hole chemistry involving higher curvature couplings \cite{Kastor:2010gq,Frassino:2014pha,Dolan:2014vba} find an array of exotic behaviors such as reentrant phase transitions and isolated critical points with unusual exponents. For general values of $N$, we expect $N$ to be a genuinely new variable and $\Omega$ to have a non-trivial dependence on $N$. In this case, our universal Smarr formula \eqref{holosmarr} no longer holds. Despite this fact \cite{Kastor:2010gq,Frassino:2014pha,Dolan:2014vba} still find generalized Smarr relations to be valid even when higher curvature couplings are included as long as one treats the coefficients of the extra curvature terms as independent couplings which can be varied as well. This appears to be an artefact of including only a finite number of curvature terms. If we have a generic function of $N$, which has an expansion at large $N$ out of which we only take the leading two (or a finite number of) terms, such that
\beq
\Omega = N^2 \Omega_0 + a N^0 \Omega_1,
\eeq
we still have a relation of the form:
\beq\label{smarr2}
 N^2 \partial_N^2 \Omega + a \partial_a \Omega = \Omega.
 \eeq
Continuing the expansion on in this manner, \eqref{smarr2} will be spoiled by further $1/N$ corrections unless one includes a new thermodynamic variable for every new term in the series. In any consistent theory of quantum gravity, such as string theory, it is believed that one has an infinite tower of higher curvature corrections, and so no useful relation of this sort will hold in general.

Maybe the simplest example where one has an interesting approximation with only two terms is the case of flavor branes \cite{Karch:2002sh}. In this case the expansion reads
\beq
\Omega = N^2 \Omega_0 + a N^1 \Omega_1 .
\eeq
These two terms are singled out by being large in the large $N$ limit and so are dominated by a semi-classical saddle. Here we have denoted $a\sim N_f$ as the number of flavors. So we can derive a generalized holographic Smarr relation in the field theory by independently varying the number of colors and flavors in the field theory. In the bulk, this corresponds to varying $G_N$ as well as the tension of the brane.

\section{Discussion}

In the preceding work, we have found that from simple field theoretic considerations alone a universal Smarr formula, \eqref{holosmarr}, emerges in holographic descriptions of black holes with large N duals. In addition, from simple scaling we were able to derive the equation of state governing the thermal system and giving necessary data to carry out our analysis. This new universal formula, when processed with the equation of state and holographic dictionary, gives bulk Smarr formulae that one would derive in the context of black hole chemistry for a wide range of distinct systems. It would be interesting to work out the consequences of our proposal in more non-trivial large $N$ field theories, such as those with Lifshitz scaling or broken global symmetries. In all these cases, our universal large $N$ Smarr formula together with the equation of state should allow one to derive a Smarr-like formula from which one can easily read of the bulk pressure as well as the thermodynamic volume.

\section*{Acknowledgements}

Thanks to Robert Mann for his stimulating talk at the Strings, Black Holes and Quantum Information Workshop at the Tohoku Forum for Creativity in Sendai, Japan, which prompted us to seek an holographic interpretation of black hole chemistry. Special thanks also to Robert Myers for helpful discussions and comments. This work is partially supported by the US Department of Energy under grant number DE-SC0011637.

\bibliographystyle{JHEP}
\bibliography{chemistry}

\end{document}